\documentclass[11pt,twoside]{article}
\usepackage{asp2010}

\resetcounters

\begin{document}

\title{Investigating the properties of granulation in the red giants observed by {\it Kepler}}
\author{S. Mathur$^{*,1}$, S. Hekker$^{2,3}$, R. Trampedach$^{4}$, J. Ballot$^{5,6}$, T. Kallinger$^{7,8}$, D. Buzasi$^{9}$, R.~A. Garc\'ia$^{10}$, D. Huber$^{11}$, A. Jim\'enez$^{12, 13}$, B. Mosser$^{14}$, T.~R. Bedding$^{11}$,  Y. Elsworth$^{3}$,  C. R\'egulo$^{12,13}$,  D. Stello$^{11}$,   W.~J. Chaplin$^{3}$, J. De Ridder$^{8}$, S.~J. Hale$^{3}$,   K. Kinemuchi$^{15}$, H. Kjeldsen$^{16}$, F. Mullaly$^{17}$ and S.~E. Thompson$^{17}$\\
{$^*$} Affiliations are given at the end of the paper}

\begin{abstract}
More than 1000 red giants have been observed by NASA/Kepler mission during a nearly continuous period of $\sim$~13 months. The resulting high-frequency resolution ($<$\,0.03~$\mu$Hz) allows us to study the granulation parameters of these stars. The granulation pattern results from the convection motions leading to upward flows of hot plasma and downward flows of cooler plasma. We fitted Harvey-like functions to the power spectra, to retrieve the timescale and amplitude of granulation. We show that there is an anti-correlation between both of these parameters and the position of maximum power of acoustic modes, while we also find a correlation with the radius, which agrees with the theory. We finally compare our results with 3D models of the convection.
\end{abstract}

\section{Analysis}

Among the targets of the {\it Kepler} mission, a large number of red giants are being followed for asteroseismic and astrometric purposes with a long-cadence sampling (29.4 min). The study of these stars is already giving very interesting results \citep[e.g.][]{2011Sci...332..205B, 2011Natur.471..608B}

We analyzed $\sim$~1000 red giants for which we could detect solar-like oscillations, see \citet{2011A&A...525A.131H}. First, the data were processed as described by \citet{2011MNRAS.414L...6G} to remove drifts, instrumental effects, and jumps. 
To remove the signature of instrumental signals with periods longer than the expected granulation time scale for red giants, we applied a triangular smooth with a width of 10 days that filters out periodicity $>$ 10 days. 

The granulation signal in these red giants was fitted by six teams \citep[][D.~L. Buzasi and B. Mosser private communications]{2009CoAst.160...74H, 2010MNRAS.402.2049H, 2010A&A...522A...1K, 2010A&A...511A..46M}. All the methods are based on the Harvey-like function \citep{1985ESASP.235..199H}:

\begin{equation}
P_{\rm H}(\nu)=\frac{4\sigma^2\tau_{\rm gran}}{1+(2\pi\nu\tau_{\rm gran})^\alpha}\ ,
\end{equation}

in which $P_{\rm H}(\nu)$ is the total granulation power of the signal at frequency $\nu$, $\sigma$ is the characteristic amplitude of the granulation and $\alpha$ is a positive parameter characterizing the slope of the decay. We also define the amplitude of the granulation power, $P_{\rm gran}~=~4\sigma^2 \tau_{\rm gran}$. 

To compare the timescales of several methods that have different values of $\alpha$, we defined the parameter $\tau_{\rm eff}$, as the e-folding time of the auto-correlation function of the temporal signal of the granulation component.

\section{Discussion}

\begin{figure}[htbp]
\begin{center}
\includegraphics[width=4cm, angle=90, trim=2cm 2cm 0 4cm]{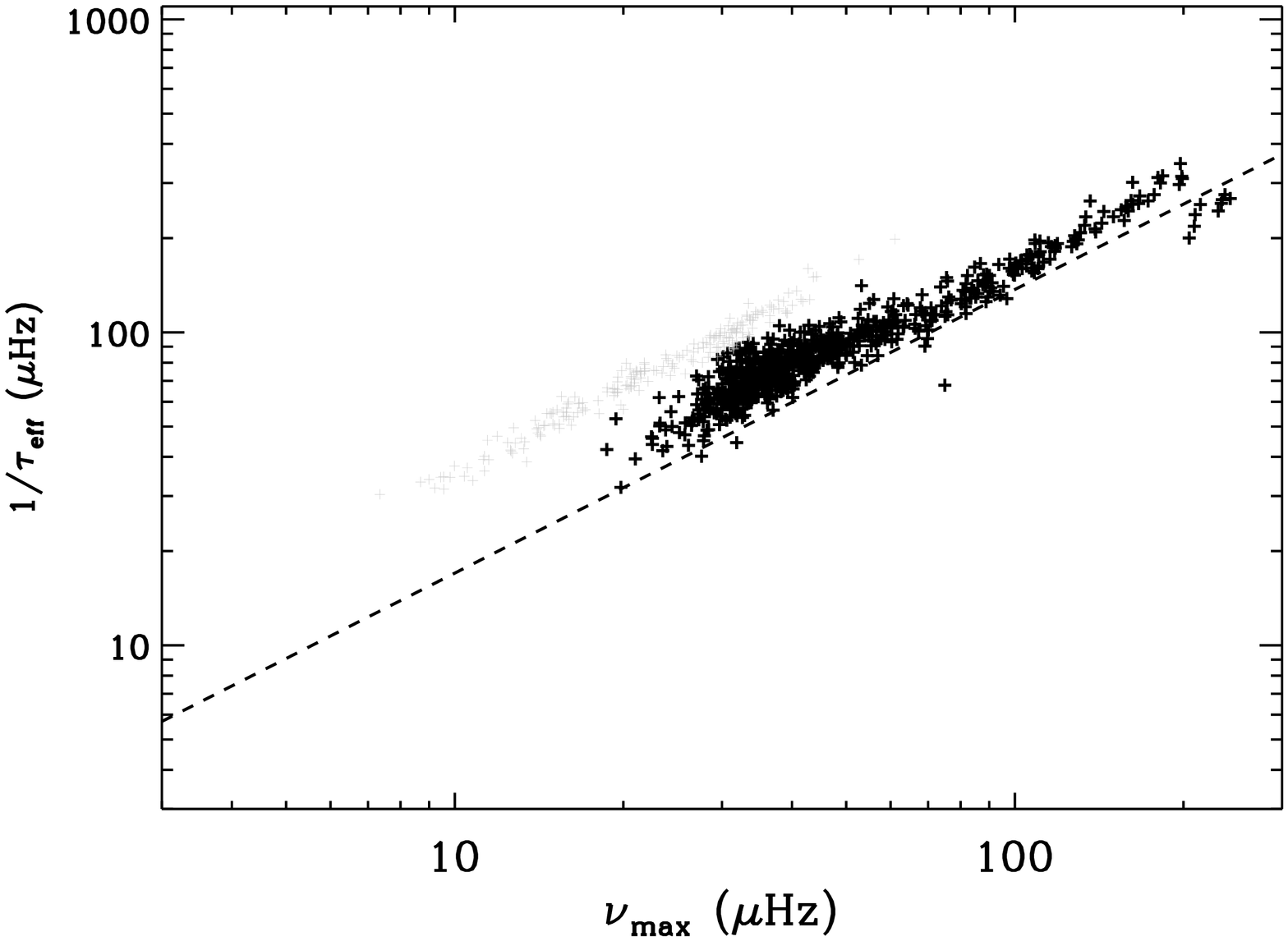}
\includegraphics[width=4cm, angle=90, trim=2cm 4cm 0 0cm]{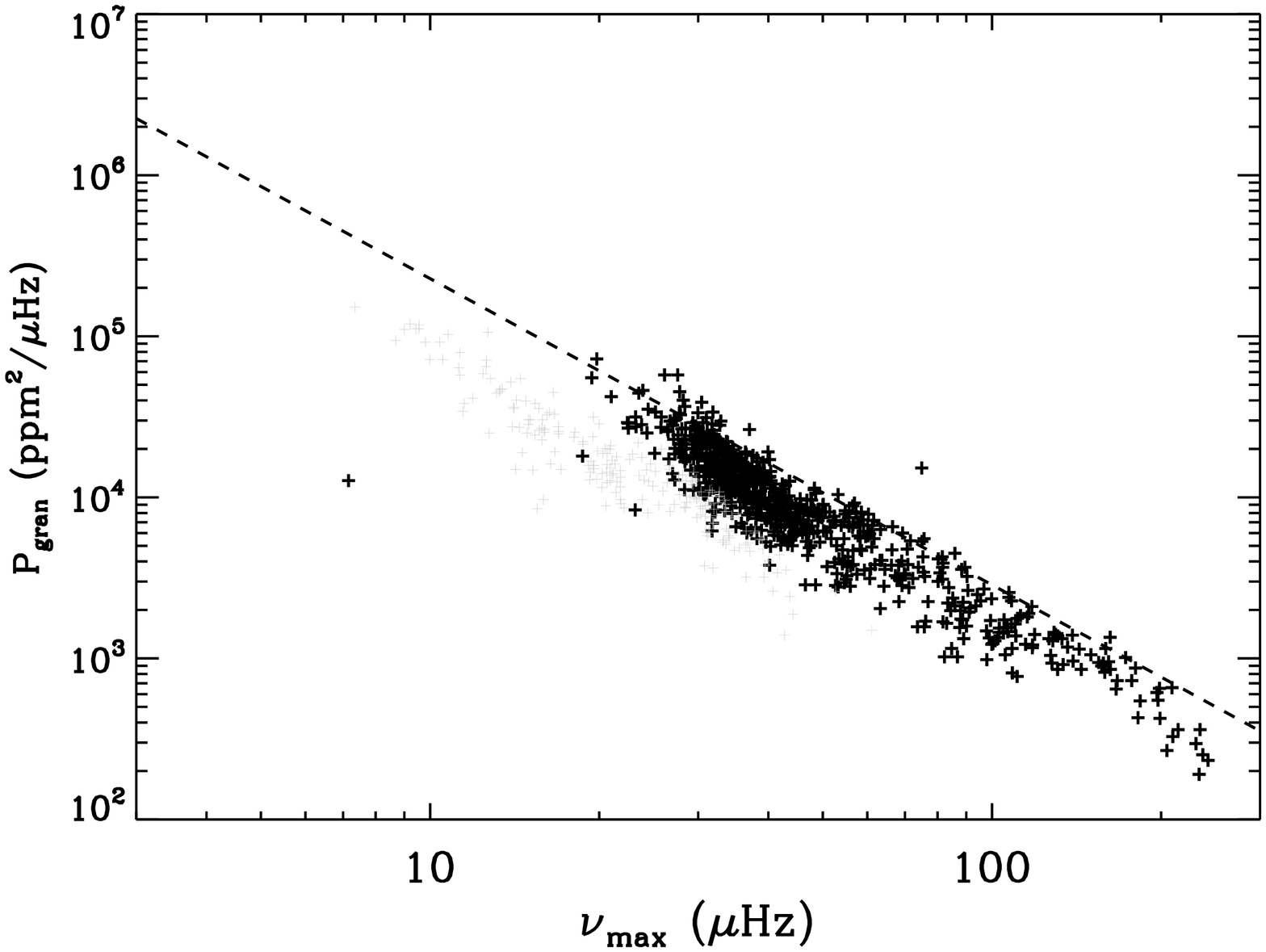}
\caption{Left panel: Characteristic granulation time scale, $\tau_{\rm eff}$, as a function of $\nu_{\rm max}$ (position of maximum power) obtained by one method  \citep{2010A&A...511A..46M}. Right panel: Granulation power ($P_{\rm gran}$) as a function of $\nu_{\rm max}$ obtained by the same method as the left panel. The light gray points correspond to a second branch picked up by the fitting code (more details are given in \citet{2011arXiv1109.1194M}). The dashed line in each panel is the result of a linear fit of $\tau_{\rm eff}$ and $P_{\rm gran}$ using the six methods all together.}
\label{fig1}
\end{center}
\end{figure}

Fig.~\ref{fig1} (left panel) shows the variation of 1/$\tau_{\rm eff}$ vs. $\nu_{\rm max}$, the frequency of maximum oscillation power for one of the six methods. Though the values can be different from one method to the other, the correlations are in agreement and we find that $\tau_{\rm eff}$ $\propto$ $\nu_{\rm max}^{-0.89}$ by taking into account all the methods together. Fig.~\ref{fig1} (right panel) shows the correlation between $P_{\rm gran}$ and $\nu_{\rm max}$ for the same method as the left panel. If we take into account the results of all the methods together, we find that $P_{\rm gran}$ $\propto$ $\nu_{\rm max}^{-1.90}$. These relations are in agreement with the scaling relations \citep{2011A&A...529L...8K}, which suggest that $\tau_{\rm gran}$ $\propto$ $\nu_{\rm max}^{-1}$ and $P_{\rm gran}$ $\propto$ $\nu_{\rm max}^{-2}$. 

Using the methodology of \citet{2010A&A...522A...1K}, we estimated the stellar parameters of 1035 red giants. 
As predicted from scaling relations for bigger stars, the granulation timescales are larger. 
We also observe an anti-correlation between $\tau_{\rm eff}$ and log $g$. 

We have computed 37 3D hydrodynamic simulations of the convection as described in \citet{2011ApJ...731...78T} and have compared them with the {\it Kepler} observations. We find similar trends but the absolute values are different by a factor of 2 for $\tau_{\rm eff}$ and by an order of magnitude for $P_{\rm gran}$.
For more details on this study, we refer to \citet{2011arXiv1109.1194M}.


\acknowledgements 
Funding for this Discovery mission is provided by NASAs Science Mission Directorate. We thank all funding councils and agencies that have supported the activities of KASC, and the International Space Science Institute (ISSI). NCAR is supported by the National Science Foundation. SH also acknowledges financial support from the Netherlands Organization for Scientific Research (NWO). JDR and TK acknowledge the support of the FWO-Flanders under project O6260 - G.0728.11.\\
Affiliations: 
{$^1$High Altitude Observatory, NCAR, P.O. Box 3000, Boulder, CO 80307, USA.}
{$^2$Astronomical Institute "Anton Pannekoek", University of Amsterdam, PO Box 94249, 1090 GE Amsterdam, The Netherlands.}
{$^3$School of Physics and Astronomy, University of Birmingham, Edgbaston, Birmingham B15 2TT, UK.}
{$^4$JILA, University of Colorado and National Institute of Standards and Technology, 440 UCB, Boulder, CO 80309, USA.}
{$^5$Institut de Recherche en Astrophysique et Plan\'etologie, Universit\'e de Toulouse, CNRS, 14 avenue E. Belin, 31400 Toulouse, France.}
{$^6$Universit\'e de Toulouse, UPS-OMP, IRAP, 31400 Toulouse, France.}
{$^7$Institute for Astronomy (IfA), University of Vienna, T\"urkenschanzstrasse 17, 1180 Vienna, Austria.}
{$^8$Instituut voor Sterrenkunde, K.U. Leuven, Celestijnenlaan 200D, 3001 Leuven, Belgium.}
{$^9$Eureka Scientific, 2452 Delmer Street Suite 100, Oakland, CA 94602-3017, USA.}
{$^{10}$La\-bo\-ra\-toi\-re AIM, CEA/DSM -- CNRS - Universit\'e Paris Diderot -- IRFU/SAp, 91191 Gif-sur-Yvette Cedex, France.}
{$^{11}$Sydney Institute for Astronomy, School of Physics, University of Sydney, NSW 2006, Australia.}
{$^{12}$Universidad de La Laguna, Dpto de Astrof\'isica, 38206, Tenerife, Spain.}
{$^{13}$Instituto de Astrof\'\i sica de Canarias, 38205, La Laguna, Tenerife, Spain.}
{$^{14}$LESIA, UMR8109, Universit\'e Pierre et Marie Curie, Universit\'e Denis Diderot, Obs. de Paris, 92195 Meudon Cedex, France.}
{$^{15}$Bay Area Environmental Research Inst./NASA Ames Research Center, Moffett Field, CA 94035, USA.}
{$^{16}$Danish AsteroSeismology Centre, Department of Physics and Astronomy, University of Aarhus, 8000 Aarhus C, Denmark.}
{$^{17}$SETI Institute/NASA Ames Research Center, Moffett Field, CA 94035, USA.}

\bibliographystyle{asp2010} 
\bibliography{/Users/Savita/Documents/BIBLIO_sav.bib}

\end{document}